\documentclass[10pt,a4paper]{article}
\usepackage{amsmath}
\usepackage{amssymb}
\usepackage{amsfonts}
\usepackage{amstext}
\usepackage{amsbsy}
\usepackage[mathscr]{eucal}
\usepackage{graphicx}
\usepackage{color}
\newcommand{\beq}{\begin{equation}}
\newcommand{\eeq}{\end{equation}}
\newcommand{\beqa}{\begin{eqnarray}}
\newcommand{\eeqa}{\end{eqnarray}}
\newcommand{\ket}[1]{| #1 \rangle}

\makeindex
\title{\Large\textbf{Complex multi-projective variety and entanglement}}

\author{\textit{Hoshang Heydari}  and \textit{Gunnar Bj\"{o}rk}\\
       \small \textit{hoshang@imit.kth.se, http://www.imit.kth.se/QEO/}\\
        \small\textit{Department of Microelectronics and}
 \small\textit{Information Technology,}\\
 \small\textit{Royal Institute of Technology (KTH)},
\small\textit{Electrum 229, SE-164 40 Kista, Sweden}}
\date{}
\begin{document}

\maketitle \thispagestyle{empty}

\maketitle
\begin{abstract}
In this paper, we will show that a vanishing  generalized
concurrence of a separable state can be seen as an algebraic
variety called the Segre variety. This variety define a quadric
space which gives a geometric picture of separable states. For
pure, bi- and three-partite states the variety equals the
generalized concurrence. Moreover, we generalize the Segre variety
to a general multipartite state by relating to a quadric space
defined by two-by-two subdeterminants.
\end{abstract}

%%%%%%%%%%%%%%%%%%%%%%%%%%%%%%%%%%%%%%%%%%%%%%%%%%
\section{Introduction}
The most interesting feature of quantum mechanical systems,
namely,  quantum entanglement, was defined by Schr\"{o}dinger
\cite{Sch35} and Einstein, Podolsky, and Rosen \cite{EPR35}. Many
years has passed since the dawn of quantum mechanics, but we have
still not been able to solve the enigma of entanglement, e.g.,
finding a complete mathematical model to describe, quantify, and
in the same time reveal the physical implications of this feature.
Moreover, we known very little  about the geometry of
entanglement. In quantum mechanics, the space of a pure state can
be described by the $N$-dimensional complex projective space
$\mathbf{CP}^{N}$. The question now is, how can we define quantum
entanglement of a general pure state on such complex projective
space?

There are several different answers to this question. One of the
earliest proposals was to quantify the entanglement in terms of a
distance to the nearest separable state \cite{Vedral97}. Another
idea is to use the maximum violation of generalized Bell
inequalities as a measure of entanglement \cite{CHSH}. Such Bell
inequality functions are called entanglement witnesses, and have
mostly been used to detect nonseparable states
\cite{Terhal,Lewenstein,Barbieri}. However, in a recent paper,
Bertlmann, Narnhofer and Thirring, have combined the two ideas and
shown that the maximal violation of a generalized Bell
inequalities and the Hilbert-Schmidt distance to the convex set of
separable states are equivalent \cite{Bertlmann}. Hence, they
demonstrate that both these concepts have a geometric
interpretation. Yet another idea to quantify entanglement is to
use the entropy of the reduced density matrix as a measure of
entanglement, the so called entanglement of formation
\cite{Bennett96}. If the entropy of the remaining subsystem is the
same as that for the original system, there is no entanglement
between the remaining subsystem and the subsystem being traced
out. For bipartite, pure states, the entanglement of formation is
simply a entropic function of the state's so-called concurrence
\cite{Wootters98}. In this paper we shall demonstrate that
concurrence, just like entanglement witnesses, has a geometric
interpretation. The connection between concurrence and geometry is
found in a map called a Segre embedding, see D. C. Brody and L. P.
Hughston \cite{Dorje99}. They illustrate this map for a pair of
qubits, and point out that this map characterizes the idea of
quantum entanglement. Moreover, they define a variety that
represents the set of separable states but they do not discuss it
much further. Segre embedding has also been discussed by A. Miyake
\cite{Miyake} in the context of classification of multipartite
states in entanglement classes (where two states belong to the
same class if they are interconvertible under stochastic local
operations and classical communication).

In this paper we will expand this idea and describe the Segre
variety, which is a quadric space in algebraic geometry, by giving
a complete and explicit formula for it. Moreover, we will compare
the Segre variety with the concurrence of a general pure,
bipartite state
\cite{Albeverio,Gerjuoy,Rungta01,Bhaktavatsala,Akhtarshenas}.
Vanishing of the concurrence of a separable state coincide with
the Segre variety. This will illustrate the geometry of
concurrence as a measure of bipartite entanglement in a complete
and satisfactory way. Furthermore, we generalize Segre variety to
a general multipartite state by relating the decomposable tensors
to a quadric space defined by two-by-two prime ideals.
%%%%%%%%%%%%%%%%%%%%%%%%%%%%%%
In this paper, we assume that the reader is familiar with basic
concepts in abstract algebra such as ring theory and fields.

%%%%%%%%%%%%%%%%%%%%%%%%%%%%%%%%%%%%%%%%%%%%%%%%%%%%%%%%%%%%%%%%
\section{Quantum entanglement}
In this section we will define separable states and entangled
states. Let us denote a general, pure, composite quantum system
with $m$ subsystems
$\mathcal{Q}=\mathcal{Q}^{p}_{m}(N_{1},N_{2},\ldots,N_{m})
=\mathcal{Q}_{1}\mathcal{Q}_{2}\cdots\mathcal{Q}_{m}$, consisting
of a state
\begin{equation}\label{Mstate}
\ket{\Psi}=\sum^{N_{1}}_{i_{1}=1}\sum^{N_{2}}_{i_{2}=1}\cdots\sum^{N_{m}}_{i_{m}=1}
\alpha_{i_{1},i_{2},\ldots,i_{m}} \ket{i_{1},i_{2},\ldots,i_{m}}
\end{equation}
 defined on a Hilbert space
\begin{eqnarray}
\mathcal{H}_{\mathcal{Q}}&=&\mathcal{H}_{\mathcal{Q}_{1}}\otimes
\mathcal{H}_{\mathcal{Q}_{2}}\otimes\cdots\otimes\mathcal{H}_{\mathcal{Q}_{m}}\\\nonumber
&=&\mathbf{C}^{N_{1}}\otimes\mathbf{C}^{N_{2}}\otimes\cdots\otimes\mathbf{C}^{N_{m}},
\end{eqnarray}
where the dimension of the $j$th Hilbert space is given  by
$N_{j}=\dim(\mathcal{H}_{\mathcal{Q}_{j}})$. We are going to use
this notation throughout this paper, i.e., we denote a pure pair
of qubits by $\mathcal{Q}^{p}_{2}(2,2)$. Next, let
$\rho_{\mathcal{Q}}$ denote a density operator acting on
$\mathcal{H}_{\mathcal{Q}}$. The density operator
$\rho_{\mathcal{Q}}$ is said to be fully separable, which we will
denote by $\rho^{sep}_{\mathcal{Q}}$, with respect to the Hilbert
space decomposition, if it can  be written as
\begin{equation}\label{eq:sep}
\rho^{sep}_{\mathcal{Q}}=\sum^\mathrm{N}_{k=1}p_k
\bigotimes^m_{j=1}\rho^k_{\mathcal{Q}_{j}},~\sum^N_{k=1}p_{k}=1
\end{equation}
 for some positive integer $\mathrm{N}$, where $p_{k}$ are positive real
numbers and $\rho^k_{\mathcal{Q}_{j}}$ denote a density operator
on Hilbert space $\mathcal{H}_{\mathcal{Q}_{j}}$. If
$\rho^{p}_{\mathcal{Q}}$ represents a pure state, then the quantum
system is fully separable if $\rho^{p}_{\mathcal{Q}}$ can be
written as
$\rho^{sep}_{\mathcal{Q}}=\bigotimes^m_{j=1}\rho_{\mathcal{Q}_{j}}$,
where $\rho_{\mathcal{Q}_{j}}$ is a density operator on
$\mathcal{H}_{\mathcal{Q}_{j}}$. If a state is not separable, then
it is called an entangled state. Some of the generic entangled
states are called Bell states and $\mathrm{EPR}$ states.

%%%%%

%&&&&&&&&&&&&&
\section{Segre Variety}
This section serves as an introduction to the affine space, Segre
embedding, and the Segre variety in such way that it enables us to
establish a relation between concurrence and Segre variety in
following sections. The general references for this section are
\cite{Li2000,Musili2001,Ueno1997,Griff78,Mum76}. Let $\mathbf{C}$
be a field of complex numbers and $N$ be an integer. Then we
define a $N$-dimensional affine space over $\mathbf{C}$, denoted
$\mathcal{A}^{N}_{\mathbf{C}}$ or $\mathcal{A}^{N}$, to be the set
of all $N$-tuples of elements of $\mathbf{C}$, i.e,
\begin{equation}
\mathcal{A}^{N}=\{P=(a_{1},a_{2},\ldots,a_{N}):a_{1},a_{2},\ldots,a_{N}\in
\mathbf{C}\}.
\end{equation}
An element $P=(a_{1},a_{2},\ldots,a_{N})$ is called a point, where
$a_{i}\in \mathbf{C}$ is called a coordinate of $P$. In general we
call $\mathcal{A}^{1}=\mathbf{C}$ the affine line and
$\mathcal{A}^{2}$ the affine plane.

Let $R(N)=\mathbf{C}[Z_{1},Z_{1},\ldots,Z_{N}]$ be the polynomial
ring over $\mathbf{C}$ in the $N$ variables
$Z_{1},Z_{1},\ldots,Z_{N}$. Any element $F\in R(N)$ gives rise to
a $\mathbf{C}$-valued map on $\mathcal{A}^{N}$ by evaluation,
i.e., $P=(a_{1},a_{2},\ldots,a_{N})\longmapsto
F(a_{1},a_{2},\ldots,a_{N})=F(P)$. Such a function on
$\mathcal{A}^{N}$ is called a polynomial or a regular function.
 Given $F\in R(n)$, the set of points yielding zeros of $F$ is denoted
 $\mathcal{V}(F)$, i.e.,
\begin{equation}
\mathcal{V}(F)=\{P\in\mathcal{A}^{N}:F(P)=0\}.
\end{equation}
A closed subset of $\mathcal{A}^{N}$ which is of the form
$\mathcal{V}(F)$, with $F\in R(N)$ not a scalar, is called the
hypersurface defined by $F$ or the hypersurface whose equation is
$F=0$. If $F\in R(N)$ is of degree $r\geq 1$, then
$\mathcal{V}(F)$ is called a hypersurface of  degree $r$ in
$\mathcal{A}^{N}$. It is called a hyperplane, a quadric, a cubic,
$\ldots$ for $r=1,2,3,\ldots$. The union of a finite number of
hypersurfaces is again a hypersurface and its degree is the sum of
their degrees, i.e.,
\begin{equation}
\mathcal{V}(F_{1}F_{2}\cdots
F_{d})=\mathcal{V}(\bigcap^{r}_{i=1}F_{i})=
\mathcal{V}(F_{1})\cup\mathcal{V}(F_{2})\cup\cdots\cup\mathcal{V}(F_{1}).
\end{equation}
A subset $\mathcal{I}$ of a commutative ring $R$ is called an
ideal of  $R$ if it has the following properties: (i) For any
elements $\alpha,\beta\in\mathcal{I}$, we have
$\alpha+\beta\in\mathcal{I}$. (ii) For any elements $a\in R$ and
$\alpha\in\mathcal{I}$, we have $a\alpha\in\mathcal{I}$. If two
elements $a\neq 0,~b\neq 0$ of $R$ satisfy $ab=0$, then we we call
$a$ a zero divisor (and so $b$).  $R$ is called an integral domain
if it has no zero divisor and an ideal $\mathcal{I}$ of $R$ is
called a prime ideal if $R/\mathcal{I}$ is an integral domain. The
ideal $\mathcal{I}(V)$ of an algebraic subset
$V\subset\mathcal{A}^{N}$ is the largest ideal of polynomial
functions on $\mathcal{A}^{N}$ vanishing on $V$ and the coordinate
ring $\mathbf{C}[V]$ of $V$ is naturally isomorphic to quotient
ring $R(N)/\mathcal{I}(V)$. $\mathbf{C}[V]$ is reduced and $V$ is
said to be equipped with the canonical reduced structure. An
irreducible algebraic subset $V$ of $\mathcal{A}^{N}$ is called an
affine algebraic variety, i.e., if its ideal $\mathcal{I}(V)$ is a
prime ideal of $R(N)$ or equivalently, its coordinate ring
$\mathbf{C}[V]=R(N)/\mathcal{I}(V)$ is an integral domain.

%%%%%%%%%%%%%%%%%%%%%%%%%
Now, let $\mathcal{A}^{N_{1}}$ and $\mathcal{A}^{N_{2}}$ be affine
spaces. If $X=(x_{1},x_{2},\ldots,x_{N_{1}})$ and
$Y=(y_{1},y_{2},\ldots,y_{N_{2}})$ are two points defined on
$\mathcal{A}^{N_{1}}$ and $\mathcal{A}^{N_{2}}$, respectively,
then the map
\begin{equation}
\begin{array}{ccc}
  \phi:\mathcal{A}^{N_{1}}\times \mathcal{A}^{N_{2}} & \longrightarrow &\mathcal{A}^{N_{1}+N_{2}} \\
 (X,Y) & \longmapsto&(x_{1},x_{2},\ldots,x_{N_{1}},y_{1},y_{2},\ldots,y_{N_{2}}) \\
\end{array}
\end{equation}
is a one-to-one and onto mapping. If $\mathcal{V}$ and
$\mathcal{U}$ are algebraic sets in $\mathcal{A}^{N_{1}}$ and
$\mathcal{A}^{N_{2}}$, respectively, then $\phi(\mathcal{V}\times
\mathcal{U})$ is a algebraic set in $\mathcal{A}^{N_{1}+N_{2}}$.

If $X=(x_{1},x_{2},\ldots,x_{N})$ and
$Y=(y_{1},y_{2},\ldots,y_{N})$ are two different points in
$\mathcal{A}^{N}$, then the line $\mathcal{L}$ passing through $X$
and $Y$ is parametrically defined as
\begin{equation}
   \mathcal{L}=\{(\delta x_{1}+ \tau y_{1},\delta x_{2}+ \tau y_{2},
   \ldots,\delta x_{N}+ \tau y_{N}):\delta,
   \tau\in\mathbf{C}\}.
\end{equation}
The complex projective space, $\mathbf{CP}^{N-1}$, is defined as
the set of all lines through $(0,0,\ldots,0)$ in
$\mathcal{A}^{N}$. Let $X$ and $Y$ be two points. Then $X$ and $Y$
determines the same line if, and only if, there exist a $\delta\in
\mathbf{C}$, $\delta\neq0$, such that $y_{i}=\delta x_{i}$, for
all $i=1,2,\ldots, N$. That is, the lines $X$ and $Y$ are
equivalent, which we denote by $X\sim Y$. Now, if we assume that
this is the case, then
\begin{equation}
\mathbf{CP}^{N-1}\cong
\frac{\mathcal{A}^{N}-\{(0,0,\ldots,0)\}}{X\sim \delta X}.
\end{equation}
If a point $X\in \mathbf{CP}^{N-1}$ is determined by
$(x_{1},x_{2},\ldots,x_{N})\in\mathcal{A}^{N}$, then we say that
$(x_{1},x_{2},\ldots,x_{N})$ is a set of homogeneous coordinates
for $X$. If $x_{i}\neq 0$, then we have
\begin{equation}
X=(\frac{x_{1}}{x_{i}},\ldots,\frac{x_{i-1}}{x_{i}},1,\frac{x_{i+1}}{x_{i}},\ldots,
\frac{x_{N}}{x_{i}}).
\end{equation}
Let $R=R(N)=\mathbf{C}[Z_{0},Z_{1},\ldots,Z_{N}]$ be  the
polynomial ring over $\mathbf{C}$ in the variables
$Z_{0},Z_{1},\ldots,Z_{N}$. Then, for a form $F\in R$, we define
$\mathcal{V}(F)=\{P\in \mathbf{CP}^{N-1}: F(P)=0\}$, called the
set of projective zeros of $F$. Unlike in the affine case, we have
$\mathbf{CP}^{N_{1}-1}\times\mathbf{CP}^{N_{2}-1}\neq
\mathbf{CP}^{N_{1}+N_{2}-2}$. For example, in
$\mathbf{CP}^{1}\times \mathbf{CP}^{1}$, the lines
$\mathcal{L}_{x}=\{x\}\times\mathbf{CP}^{1}$ and
$\mathcal{L}_{y}=\{y\}\times\mathbf{CP}^{1}$ are parallel for
$x\neq y$ in $\mathbf{CP}^{1}$ but there are no parallel lines in
$\mathbf{CP}^{2}$ since any two distinct lines
$L_{1}=\mathcal{V}(a_{1}X_{1}+a_{2}X_{2}+a_{3}X_{3})$ and
$L_{2}=\mathcal{V}(b_{1}X_{1}+b_{2}X_{2}+b_{3}X_{3})$ intersect at
the unique point
$(a_{2}b_{3}-a_{3}b_{2},a_{3}b_{1}-a_{1}b_{3},a_{1}b_{2}-a_{2}b_{1})$.

 Now, we want to make $\mathbf{CP}^{N_{1}-1}\times\mathbf{CP}^{N_{2}-1}$
into a projective variety by its Segre embeding which we construct
as follows: Let $X$ and $Y$ be two points defined on
$\mathbf{CP}^{N_{1}-1}$ and $\mathbf{CP}^{N_{2}-1}$, respectively.
Then the map
\begin{equation}
\begin{array}{ccc}
  \mathcal{S}_{N_{1},N_{2}}:\mathbf{CP}^{N_{1}-1}\times\mathbf{CP}^{N_{2}-1}&\longrightarrow&
\mathbf{CP}^{N_{1}N_{2}-1}\\
 (X,Y) & \longmapsto&(x_{1}y_{1},\ldots,x_{1}y_{N_{2}},\ldots,x_{N_{1}}y_{1},\ldots,x_{N_{1}}y_{N_{2}}) \\
\end{array}
\end{equation}
is a closed immersion, called the Segre embedding. To see that,
let $X_{i}$, and $Y_{j}$ be the homogeneous coordinate functions
on $\mathbf{CP}^{N_{1}-1}$ and $\mathbf{CP}^{N_{2}-1}$,
respectively. Moreover, let $Z_{i,j}$ be the homogeneous
coordinate-function on $\mathbf{CP}^{N_{1}N_{2}-1}$. Now, we
arrange the homogeneous coordinate $Z_{i,j}$ as follows
\begin{equation}
\left(%
\begin{array}{cccc}
  Z_{1,1} & Z_{1,2} & \cdots & Z_{1,N_{2}} \\
  Z_{2,1} & Z_{2,2} & \cdots & Z_{2,N_{2}} \\
  \vdots & \vdots&\ddots &\vdots \\
   Z_{N_{1},1} & Z_{N_{1},2} & \cdots & Z_{N_{1},N_{2}} \\
\end{array}%
\right)
\end{equation}
The map $\mathcal{S}_{N_{1},N_{2}}=(\cdots, X_{i}Y_{j},\cdots)$ is
a morphism since it is defined by polynomials on any affine piece
$U_{i}\times U_{j}$ where
\begin{equation}
\mathbf{CP}^{N_{1}-1}=\bigcup^{N_{1}-1}_{i=1}U_{i} ~\text{and}
  ~\mathbf{CP}^{N_{2}-1}=\bigcup^{N_{2}-1}_{j=1}U_{j}
\end{equation}
are the standard affine coverings.
But the determinant
\begin{equation}
\det\left(%
\begin{array}{cc}
  X_{i}Y_{k} & X_{i}Y_{l} \\
  X_{j}Y_{k} & X_{j}Y_{l} \\
\end{array}%
\right)
\end{equation}
vanishes for all $i,j$ and $k,l$, so the image of
$\mathcal{S}_{N_{1},N_{2}}$ is contained in the closed subset
\begin{eqnarray}
T&=&\nonumber\left\{\right(\cdots,
z_{i,j},\cdots)\in\mathbf{CP}^{N_{1}N_{2}-1}
:\mathrm{rk}\left(%
\begin{array}{cccc}
  z_{1,1} & z_{1,2} & \cdots & z_{1,N_{2}} \\
  z_{2,1} & z_{2,2} & \cdots & z_{2,N_{2}} \\
  \vdots & \vdots&\ddots &\vdots \\
   z_{N_{1},1} & z_{N_{1},2} & \cdots & z_{N_{1},N_{2}} \\
\end{array}%
\right)=1\},
\end{eqnarray}
where $\mathrm{rk}$ denotes the matrix rank. If $\mathrm{Im}$
denotes the image, then
$T=\mathrm{Im}\left(\mathcal{S}_{N_{1},N_{2}}\right)$ and
$\mathcal{S}_{N_{1},N_{2}}$ is an isomorphism. To see that, let us
consider
 $t=(\cdots, z_{i,j},\cdots)\in Z$. Then all the rows and
columns of the rank one matrix $(z_{i,j})$ are proportional. For
any columns $x\neq0$ and any rows $y\neq 0$ of this matrix we have
$t=\mathcal{S}_{N_{1},N_{2}}(x,y)$ and
$T=\mathrm{Im}(\mathcal{S}_{N_{1},N_{2}})$. Moreover, the map
$t\longmapsto(x,y)$ is the inverse to $\mathcal{S}_{N_{1},N_{2}}$
and so it is an isomorphism. If $V\subseteq\mathbf{CP}^{N_{1}-1}$
and $W\subseteq\mathbf{CP}^{N_{2}-1}$ are projective algebraic
sets, then $V\times W$ is projective and is closed in the closed
subvariety
$\mathbf{CP}^{N_{1}-1}\times\mathbf{CP}^{N_{2}-1}=\mathrm{Im}(\mathcal{S}_{N_{1},N_{2}})
\subset\mathbf{CP}^{N_{1}N_{2}-1}$. The image of the Segre
embedding is an intersection of a family of quadric hypersurfaces
in $\mathbf{CP}^{N_{1}N_{2}-1}$, that is
\begin{eqnarray}
\mathrm{Im}(\mathcal{S}_{N_{1},N_{2}})&=&
\bigcap_{i,j,k,l}\mathcal{V}\left(
  Z_{i,k}Z_{j,l}-Z_{i,l} Z_{j,k}\right).
\end{eqnarray}
I.e.,
$\mathrm{Im}(\mathcal{S}_{2,2})=\mathcal{V}\left(Z_{1,1}Z_{2,2}-Z_{1,2}
Z_{2,1}\right)$ is a quadric surface in $\mathbf{CP}^{3}$.
\subsection{Segre variety for a general bipartite state and concurrence}

For given quantum system $\mathcal{Q}_{2}(N_{1},N_{2})$ we want
make $\mathbf{CP}^{N_{1}-1}\times\mathbf{CP}^{N_{2}-1}$ into a
projective variety by its Segre embedding which we construct as
follows. Let $(\alpha_{1},\alpha_{2},\ldots,\alpha_{N_{1}})$ and
$(\alpha_{1},\alpha_{2},\ldots,\alpha_{N_{2}})$ be two points
defined on $\mathbf{CP}^{N_{1}-1}$ and $\mathbf{CP}^{N_{2}-1}$,
respectively, then the Segre map
\begin{equation}
\begin{array}{cc}
  \mathcal{S}_{N_{1},N_{2}}:\mathbf{CP}^{N_{1}-1}
  \times\mathbf{CP}^{N_{2}-1}&\longrightarrow \mathbf{CP}^{N_{1}N_{2}-1}\\
\end{array}
\end{equation}
\begin{equation}
\begin{array}{c}
((\alpha_{1},\alpha_{2},\ldots,\alpha_{N_{1}}),
 (\alpha_{1},\alpha_{2},\ldots,\alpha_{N_{2}}))\longmapsto\\
 (\alpha_{1,1},\alpha_{1,2},\ldots,\alpha_{1,N_{1}},\ldots,\alpha_{N_{1},1},\ldots,\alpha_{N_{1},N_{2}})\\
\end{array}
\end{equation}
 is well defined. Next, let $\alpha_{i,j}$ be the homogeneous coordinate function
on $\mathbf{CP}^{N_{1}N_{2}-1}$. Then the image of the Segre
embedding is an intersection of a family of quadric hypersurfaces
in $\mathbf{CP}^{N_{1}N_{2}-1}$, that is
\begin{eqnarray}
\mathrm{Im}(\mathcal{S}_{N_{1},N_{2}})
&=&\bigcap_{i,j,k,l}\mathcal{V}\left(\mathcal{C}_{i,j;k,l}(N_{1},N_{2})\right)
 \\\nonumber&=&
 \bigcap_{i,j,k,l}\mathcal{V}\left(
  \alpha_{i,k} \alpha_{j,l}-\alpha_{i,l}  \alpha_{j,k}
 \right).
\end{eqnarray}
This quadric space is the space of separable states and it
coincides with the definition of general concurrence
$\mathcal{C}(\mathcal{Q}_{2}(N_{1},N_{2}))$ of a pure bipartite
state \cite{Albeverio,Gerjuoy} because

\begin{eqnarray}\label{Conc}
\mathcal{C}(\mathcal{Q}_{2}(N_{1},N_{2}))&=&\left(\mathcal{N}\sum^{N_{1}}_{j,i=1}\sum^{N_{2}}_{l,k=1}
\left|\mathcal{C}_{i,j;k,l}(N_{1},N_{2})\right|^{2}\right)^{\frac{1}{2}}
\\\nonumber&=&
\left(\mathcal{N}\sum^{N_{1}}_{j,i=1}\sum^{N_{2}}_{l,k=1}
\left|\alpha_{i,k}\alpha_{j,l}-
\alpha_{i,l}\alpha_{j,k}\right|^{2}\right)^{\frac{1}{2}},
\end{eqnarray}
where $\mathcal{N}$ is a somewhat arbitrary normalization
constant. The separable set is defined by
$\alpha_{i,k}\alpha_{j,l}= \alpha_{il}\alpha_{jk}$ for all $i,j$
and $k,l$. I.e.,
\begin{equation}
\mathrm{Im}(\mathcal{S}_{2,2})=\mathcal{V}
\left(\alpha_{1,1}\alpha_{2,2}-\alpha_{1,2}\alpha_{2,1}\right)
\Longleftrightarrow\alpha_{1,1}\alpha_{2,2}=\alpha_{1,2}\alpha_{2,1}
\end{equation}
 is a
quadric surface in $\mathbf{CP}^{3}$ which coincides with the
space of separable set of pairs of qubits.

\section{Multi-projective  variety and multi-partite entanglement
measure} In this section, we will generalize the Segre variety to
a multi-projective space. As in the previous section, we can make
$\mathbf{CP}^{N_{1}-1}\times\mathbf{CP}^{N_{2}-1}
\times\cdots\times\mathbf{CP}^{N_{m}-1}$ into a projective variety
by its Segre embedding following almost the same procedure. Let
$(\alpha_{1},\alpha_{2},\ldots,\alpha_{N_{i}})$  be points defined
on $\mathbf{CP}^{N_{i}-1}$. Then the Segre map
\begin{equation}
\begin{array}{ccc}
  \mathcal{S}_{N_{1},\ldots,N_{m}}:\mathbf{CP}^{N_{1}-1}\times\mathbf{CP}^{N_{2}-1}
\times\cdots\times\mathbf{CP}^{N_{m}-1}&\longrightarrow&
\mathbf{CP}^{N_{1}N_{2}\cdots N_{m}-1}\\
 ((\alpha_{1},\alpha_{2},\ldots,\alpha_{N_{1}}),\ldots,
 (\alpha_{1},\alpha_{2},\ldots,\alpha_{N_{m}})) & \longmapsto&
 (\ldots,\alpha_{i_{1},i_{2},\ldots, i_{m}},\ldots). \\
\end{array}
\end{equation}
is well defined for $\alpha_{i_{1}i_{2}\cdots i_{m}}$,$1\leq
i_{1}\leq N_{1}, 1\leq i_{2}\leq N_{2},\ldots, 1\leq i_{m}\leq
N_{m}$ as a homogeneous coordinate-function on
$\mathbf{CP}^{N_{1}N_{2}\cdots N_{m}-1}$. Now, let us consider the
composite quantum system
$\mathcal{Q}^{p}_{m}(N_{1},N_{2},\ldots,N_{m})$ and let the
coefficients of $\ket{\Psi}$, namely
$\alpha_{i_{1},i_{2},\ldots,i_{m}}$, make an array as follows
\begin{equation}
\mathcal{A}=\left(\alpha_{i_{1},i_{2},\ldots,i_{m}}\right)_{1\leq
i_{j}\leq N_{j}},
\end{equation}
for all $j=1,2,\ldots,m$. $\mathcal{A}$ can be realized as the
following set $\{(i_{1},i_{2},\ldots,i_{m}):1\leq i_{j}\leq
N_{j},\forall~j\}$, in which each point
$(i_{1},i_{2},\ldots,i_{m})$ is assigned the value
$\alpha_{i_{1},i_{2},\ldots,i_{m}}$. Then $\mathcal{A}$ and it's
realization is called an $m$-dimensional box-shape matrix of size
$N_{1}\times N_{2}\times\cdots\times N_{m}$, where we  associate
to each such matrix a sub-ring
$\mathrm{S}_{\mathcal{A}}=\mathbf{C}[\mathcal{A}]\subset\mathrm{S}$,
where $\mathrm{S}$ is a commutative ring over the complex number
field. For each $j=1,2,\ldots,m$, a two-by-two minor about the
$j$-th coordinate of $\mathcal{A}$ is given by
\begin{eqnarray}
\mathcal{C}_{k_{1},l_{1};k_{2},l_{2};\ldots;k_{m},l_{m}}&=&
\alpha_{k_{1},k_{2},\ldots,k_{m}}\alpha_{l_{1},l_{2},\ldots,l_{m}}
\\\nonumber&&-
\alpha_{k_{1},k_{2},\ldots,k_{j-1},l_{j},k_{j+1},\ldots,k_{m}}\alpha_{l_{1},l_{2},
\ldots,l_{j-1},k_{j},l_{j+1},\ldots,l_{m}}\in
\mathrm{S}_{\mathcal{A}}.
\end{eqnarray}
Then the ideal $\mathcal{I}^{m}_{\mathcal{A}}$ of
$\mathrm{S}_{\mathcal{A}}$ is generated by
$\mathcal{C}_{k_{1},l_{1};k_{2},l_{2};\ldots;k_{m},l_{m}}$  and
describes the separable states in $\mathbf{CP}^{N_{1}N_{2}\cdots
N_{m}-1}$ \cite{Grone}. The image of the Segre embedding
$\mathrm{Im}(\mathcal{S}_{N_{1},N_{2},\ldots,N_{m}})$ which again
is an intersection of families of quadric hypersurfaces in
$\mathbf{CP}^{N_{1}N_{2}\cdots N_{m}-1}$ is given by
\begin{eqnarray}\label{eq: submeasure}
\mathrm{Im}(\mathcal{S}_{N_{1},N_{2},\ldots,N_{m}})&=&\bigcap_{\forall
j}\mathcal{I}^{m}_{\mathcal{A}}\\\nonumber &=&\bigcap_{\forall
j}\mathcal{V}\left(\mathcal{C}_{k_{1},l_{1};k_{2},l_{2};\ldots;k_{m},l_{m}}\right).
\end{eqnarray}
Moreover, following the same argumentation as in bipartite case,
we can define an entanglement measure for a pure multipartite
state as
\begin{eqnarray}\label{EntangSeg}
\mathcal{E}(\mathcal{Q}^{p}_{m}(N_{1},\ldots,N_{m}))&=&\left(\mathcal{N}\sum_{\forall
j}\left|\mathcal{C}_{k_{1},l_{1};k_{2},l_{2};\ldots;k_{m},l_{m}}\right|^{2}\right)^{\frac{1}{2}}
\\\nonumber
&=&(\mathcal{N}\sum_{\forall
j}|\alpha_{k_{1},k_{2},\ldots,k_{m}}\alpha_{l_{1},l_{2},\ldots,l_{m}}
\\\nonumber&&-
\alpha_{k_{1},k_{2},\ldots,k_{j-1},l_{j},k_{j+1},\ldots,k_{m}}\alpha_{l_{1},l_{2},
\ldots,l_{j-1},k_{j},l_{j+1},\ldots,l_{m}}|^{2})^{\frac{1}{2}},
\end{eqnarray}
where $\mathcal{N}$ is an arbitrary normalization constant and
$j=1,2,\ldots,m$. This measure coincide with the concurrence for a
general bipartite and three-partite state. However, for reasons
that will be explained below, it fails to quantify the
entanglement for $m\geq 4$, whereas it still provides the
condition of full separability.
%%%%%%%%%%%%%%%%%%%%%%%%%%%%%%%
%%%%%%%%%%%%%%%%%%%%%%%%%%%%%
\section{Example: Three-partite state}
 As an example, let us look a general
three-partite state. The generalized concurrence \cite{Albeverio}
for such a state is given by
%%%%%%%%%%%%%%%%%%%%%%%%%%%%
\begin{eqnarray}
\mathcal{E}(\mathcal{Q}^{p}_{3}(N_{1},N_{2},N_{3}))&=&\left(\mathcal{N}\sum_{k_{1},l_{1};k_{2},l_{2};k_{3},l_{3}}
\sum_{\forall
j}\left|\mathcal{C}_{k_{1},l_{1};k_{2},l_{2};k_{3},l_{3}}\right|^{2}\right)^{\frac{1}{2}}\\\nonumber
&=&(\mathcal{N}\sum_{k_{1},l_{1};k_{2},l_{2};k_{3},l_{3}}
(\left|\alpha_{k_{1},k_{2},k_{3}}\alpha_{l_{1},l_{2},l_{3}}-
\alpha_{k_{1},k_{2},l_{3}}\alpha_{l_{1},l_{2},k_{3}}\right|^{2}\\\nonumber&&
+\left|\alpha_{k_{1},k_{2},k_{3}}\alpha_{l_{1},l_{2},l_{3}}-
\alpha_{k_{1},l_{2},k_{3}}\alpha_{l_{1},k_{2},l_{3}}\right|^{2})\\\nonumber&&+
\left|\alpha_{k_{1},k_{2},k_{3}}\alpha_{l_{1},l_{2},l_{3}}-
\alpha_{l_{1},k_{2},k_{3}}\alpha_{k_{1},l_{2},l_{3}}\right|^{2})^{\frac{1}{2}}.
\end{eqnarray}
This equation for an entanglement measure is equivalent but not
equal to our entanglement tensor based on  joint POVMs on phase
space \cite{Hosh4}.
%%%%%%%%%%%%%%%%%%%%%%%%%%%%%%%%%%%%%%%%%%%%%%%%%%%%
For a three-qubit state $\mathcal{Q}^{p}_{3}(2,2,2)$, we have
%%%%%%%%%%%%%%%%%%%%%%%%%%%%%%%%%%%%
\begin{eqnarray}
\mathcal{E}(\mathcal{Q}^{p}_{3}(2,2,2))&=&(4\mathcal{N}\{2 |
\alpha_{1,1,1}\alpha_{2,2,1}-\alpha_{1,2,1}\alpha_{2,1,1}
|^{2}\\\nonumber&&
+2|\alpha_{1,1,2}\alpha_{2,2,2}-\alpha_{1,2,2}\alpha_{2,1,2}|^{2}\\\nonumber&&+
2|
\alpha_{1,1,1}\alpha_{2,1,2}-\alpha_{1,1,2}\alpha_{2,1,1}|^{2}\\\nonumber&&
 +2|\alpha_{1,2,1}\alpha_{2,2,2}-\alpha_{1,2,2}\alpha_{2,2,1}|^{2}\\\nonumber&&+
 2|\alpha_{1,1,1}\alpha_{1,2,2}-\alpha_{1,1,2}\alpha_{1,2,1}|^{2}\\\nonumber&&+
2|\alpha_{2,1,1}\alpha_{2,2,2}-\alpha_{2,1,2}\alpha_{2,2,1}|^{2}\\\nonumber&&
+|
\alpha_{1,1,1}\alpha_{2,2,2}-\alpha_{1,1,2}\alpha_{2,2,1}|^{2}\\\nonumber&&+
|\alpha_{1,1,1}\alpha_{2,2,2}- \alpha_{1,2,1}\alpha_{2,1,2} |^{2}
%%%%%%%%%%%%
\\\nonumber&&+
| \alpha_{1,1,1}\alpha_{2,2,2}- \alpha_{1,2,2}\alpha_{2,1,1}
|^{2}\\\nonumber&& + |\alpha_{1,1,2}\alpha_{2,2,1}-
\alpha_{1,2,1}\alpha_{2,1,2}|^{2}\\\nonumber&&
 %%%%%%%%%%%%
+ | \alpha_{1,1,2}\alpha_{2,2,1}- \alpha_{1,2,2}\alpha_{2,1,1}
|^{2}\\\nonumber&& + |\alpha_{1,2,1}\alpha_{2,1,2}-
\alpha_{1,2,2}\alpha_{2,1,1}
 |^{2}\})^{\frac{1}{2}}.
\end{eqnarray}
%%%%%%%%%%%%%%%%%%%%%%%%%%%%%%%%%%%%%%
We can derive this expression in a different way than it was
originally derived using the idea of the Segre ideal.  The ideal
$\mathcal{I}^{2,2,2}_{\mathcal{Q}_{1}\models\mathcal{Q}_{2}\mathcal{Q}_{3}}$
representing if a subsystem $\mathcal{Q}_{1}$ that is unentangled
with  $\mathcal{Q}_{2}\mathcal{Q}_{3}$ is generated by the six
2-by-2 subdeterminants of
\begin{equation}
\left(%
\begin{array}{cccc}
  \alpha_{1,1,1} & \alpha_{1,1,2}&\alpha_{1,2,1}&\alpha_{1,2,2} \\
 \alpha_{2,1,1} & \alpha_{2,1,2}&\alpha_{2,2,1}&\alpha_{2,2,2} \\
\end{array}%
\right)
\end{equation}
and is given by
\begin{eqnarray}
\mathcal{I}^{2,2,2}_{\mathcal{Q}_{1}\models\mathcal{Q}_{2}\mathcal{Q}_{3}}&=&\nonumber
\langle\alpha_{1,1,1}\alpha_{2,1,2}-\alpha_{1,1,2}\alpha_{2,1,1},
\alpha_{1,1,1}\alpha_{2,2,1}-\alpha_{1,2,1}\alpha_{2,1,1}\\\nonumber&&
%%%%%%%%%%%%%%%%%%%%%%%%%%%%%%%%%%%
,\alpha_{1,1,1}\alpha_{2,2,2}-\alpha_{1,2,2}\alpha_{2,1,1},
\alpha_{1,1,2}\alpha_{2,2,1}-\alpha_{1,2,1}\alpha_{2,1,2}
\\\nonumber&&
%%%%%%%%%%%%%%%%%%%%%%%%%%%%%%%%%%%
,\alpha_{1,1,2}\alpha_{2,2,2}-\alpha_{1,2,2}\alpha_{2,1,2},
\alpha_{1,2,1}\alpha_{2,2,2}-\alpha_{1,2,2}\alpha_{2,2,1}\rangle,
\end{eqnarray}
%%%%%%%%%%%%%%%%%%%%%%%%%%%%%%%%%%%%%
where we have used the notation $\models$ to indicate that
$\mathcal{Q}_{1}$ is separated from
$\mathcal{Q}_{2}\mathcal{Q}_{3}$ but we still could have
entanglement between $\mathcal{Q}_{2}$ and $\mathcal{Q}_{3}$. The
notation $\{2,2,2\}$ is used to indicate a three-partite state
where the dimension of the Hilbert space of each subsystem is $2$
(i.e., three qubits).
%%%%%%%%%%%%%%%%%%%%%%%%%%%%%%%%%%%%%
In the same way, we can define the ideal
$\mathcal{I}^{2,2,2}_{\mathcal{Q}_{2}\models\mathcal{Q}_{1}\mathcal{Q}_{3}}$
representing if the subsystem $\mathcal{Q}_{2}$ is unentangled
with $\mathcal{Q}_{1}\mathcal{Q}_{3}$ and
$\mathcal{I}_{\mathcal{Q}_{3}\models\mathcal{Q}_{1}\mathcal{Q}_{2}}$
representing if the subsystem $\mathcal{Q}_{3}$ is unentangled
with $\mathcal{Q}_{2}\mathcal{Q}_{3}$. The ideals are generated by
the six 2-by-2 subdeterminants of
\begin{equation}
\left(%
\begin{array}{cccc}
  \alpha_{1,1,1} & \alpha_{1,1,2}&\alpha_{2,1,1}&\alpha_{2,1,2} \\
 \alpha_{1,2,1} & \alpha_{1,2,2}&\alpha_{2,2,1}&\alpha_{2,2,2} \\
\end{array}%
\right)~   \text{and} ~\left(%
\begin{array}{cccc}
  \alpha_{1,1,1} & \alpha_{1,2,1}&\alpha_{2,1,1}&\alpha_{2,2,1} \\
 \alpha_{1,1,2} & \alpha_{1,2,2}&\alpha_{2,1,2}&\alpha_{2,2,2} \\
\end{array}%
\right),
\end{equation}
respectively. Written out explicitly they are
\begin{eqnarray}
\mathcal{I}^{2,2,2}_{\mathcal{Q}_{2}\models\mathcal{Q}_{1}\mathcal{Q}_{3}}&=&\nonumber
\langle\alpha_{1,1,1}\alpha_{1,2,2}-\alpha_{1,1,2}\alpha_{1,2,1},
\alpha_{1,1,1}\alpha_{2,2,1}-\alpha_{2,1,1}\alpha_{1,2,1}\\\nonumber&&
%%%%%%%%%%%%%%%%%%%%%%%%%%%%%%%%%%%
,\alpha_{1,1,1}\alpha_{2,2,2}-\alpha_{2,1,2}\alpha_{1,2,1},
\alpha_{1,1,2}\alpha_{2,2,1}-\alpha_{2,1,1}\alpha_{1,2,2}
\\\nonumber&&
%%%%%%%%%%%%%%%%%%%%%%%%%%%%%%%%%%%
,\alpha_{1,1,2}\alpha_{2,2,2}-\alpha_{1,2,2}\alpha_{2,1,2},
\alpha_{2,1,1}\alpha_{2,2,2}-\alpha_{2,1,2}\alpha_{2,2,1}\rangle,
\end{eqnarray}
and
\begin{eqnarray}
\mathcal{I}^{2,2,2}_{\mathcal{Q}_{3}\models\mathcal{Q}_{1}\mathcal{Q}_{2}}&=&\nonumber
\langle\alpha_{1,1,1}\alpha_{1,2,2}-\alpha_{1,2,1}\alpha_{1,1,2},
\alpha_{1,1,1}\alpha_{2,1,2}-\alpha_{2,1,1}\alpha_{1,1,2}\\\nonumber&&
%%%%%%%%%%%%%%%%%%%%%%%%%%%%%%%%%%%
,\alpha_{1,1,1}\alpha_{2,2,2}-\alpha_{2,2,1}\alpha_{1,1,2},
\alpha_{1,2,1}\alpha_{2,1,2}-\alpha_{2,1,1}\alpha_{1,2,2}
\\\nonumber&&
%%%%%%%%%%%%%%%%%%%%%%%%%%%%%%%%%%%
,\alpha_{1,2,1}\alpha_{2,2,2}-\alpha_{2,2,1}\alpha_{1,2,2},
\alpha_{2,1,1}\alpha_{2,2,2}-\alpha_{2,2,1}\alpha_{2,1,2}\rangle.
\end{eqnarray}
Hence, the Segre ideal of a completely separable pure three-qubit
state is given by
\begin{eqnarray}
\mathcal{I}^{2,2,2}_{Segre}&=&
\mathcal{I}^{2,2,2}_{\{\mathcal{Q}_{1}\models\mathcal{Q}_{2}\mathcal{Q}_{3},
\mathcal{Q}_{2}\models\mathcal{Q}_{1}\mathcal{Q}_{3}
,\mathcal{Q}_{3}\models\mathcal{Q}_{1}\mathcal{Q}_{2}\}}\\\nonumber
&=&
\mathcal{I}^{2,2,2}_{\mathcal{Q}_{1}\models\mathcal{Q}_{2}\mathcal{Q}_{3}}\bigcap
\mathcal{I}^{2,2,2}_{\mathcal{Q}_{2}\models\mathcal{Q}_{1}\mathcal{Q}_{3}}
\bigcap\mathcal{I}^{2,2,2}_{\mathcal{Q}_{2}\models\mathcal{Q}_{1}\mathcal{Q}_{2}}
%%%%%%%%%%%
\\\nonumber&=&
\langle\alpha_{1,1,1}\alpha_{2,1,2}-\alpha_{1,1,2}\alpha_{2,1,1},
\alpha_{1,1,1}\alpha_{2,2,1}-\alpha_{1,2,1}\alpha_{2,1,1}\\\nonumber&&
%%%%%%%%%%%%%%%%%%%%%%%%%%%%%%%%%%%
,\alpha_{1,1,1}\alpha_{2,2,2}-\alpha_{1,2,2}\alpha_{2,1,1},
\alpha_{1,1,2}\alpha_{2,2,1}-\alpha_{1,2,1}\alpha_{2,1,2}
\\\nonumber&&
%%%%%%%%%%%%%%%%%%%%%%%%%%%%%%%%%%%
,\alpha_{1,1,2}\alpha_{2,2,2}-\alpha_{1,2,2}\alpha_{2,1,2},
\alpha_{1,2,1}\alpha_{2,2,2}-\alpha_{1,2,2}\alpha_{2,2,1}
\\\nonumber&&
%%%%%%%%%%%%%%%%%%%%%%%%%%%%%%%%%%%
,\alpha_{1,1,1}\alpha_{1,2,2}-\alpha_{1,1,2}\alpha_{1,2,1},
\alpha_{1,1,1}\alpha_{2,2,2}-\alpha_{1,2,1}\alpha_{2,1,2}\\\nonumber&&
%%%%%%%%%%%%%%%%%%%%%%%%%%%%%%%%%%%
,\alpha_{1,1,2}\alpha_{2,2,1}-\alpha_{1,2,2}\alpha_{2,1,1}, ,
\alpha_{2,1,1}\alpha_{2,2,2}-\alpha_{2,1,2}\alpha_{2,2,1}
\\\nonumber&&
%%%%%%%%%%%%%%%%%%%%%%%%%%%%%%%%%%%
,\alpha_{1,1,1}\alpha_{2,2,2}-\alpha_{1,1,2}\alpha_{2,2,1}
,\alpha_{1,2,1}\alpha_{2,1,2}-\alpha_{1,2,2}\alpha_{2,1,1}
\rangle.
\end{eqnarray}
This equation coincide with Eqn. (\ref{eq: submeasure}) for a
three-qubit state.
 %%%%%%%%%%%%%%%%%%%%%%%%%%%
For a general multipartite state, that is, for $m\geq 4$ this
measure $\mathcal{E}(\mathcal{Q}^{p}_{m}(N_{1},\ldots,N_{m}))$ is
not invariant under local operations. To show why this measure is
not invariant under local operations, let us consider the quantum
system $\mathcal{Q}^{p}_{4}(2,2,2,2)$. In this case we can have
seven types of separability between different subsystems as
follows: It may be possible to factor $\mathcal{Q}_{1}$,
$\mathcal{Q}_{2}$,
 $\mathcal{Q}_{3}$, or
 $\mathcal{Q}_{4}$ from the composite system. To check this we need to
make four different permutation of indices and it is exactly what
the measure $\mathcal{E}(\mathcal{Q}^{p}_{4}(2,2,2,2))$ does. But
there  is other types of separability in this four-qubit state,
namely if it is possible to factor out
$\mathcal{Q}_{1}\mathcal{Q}_{2}$,
$\mathcal{Q}_{1}\mathcal{Q}_{3}$,
$\mathcal{Q}_{1}\mathcal{Q}_{4}$,
$\mathcal{Q}_{2}\mathcal{Q}_{3}$,
$\mathcal{Q}_{2}\mathcal{Q}_{4}$, or
$\mathcal{Q}_{3}\mathcal{Q}_{4}$. These six possible
factorizations can be reduced to three checks of separability
since if we test for separability of, i.e.,
$\mathcal{Q}_{1}\mathcal{Q}_{2}$, we have simultaneously tested
$\mathcal{Q}_{3}\mathcal{Q}_{4}$. For these types of separability
we do need to perform more than one simultaneous permutation of
indices. The measure (\ref{EntangSeg}) does not check this type of
separability which is needed in general case \cite{Pan}.

%%%%%%%%%%%%%%%%%%%%%%%%%%%%%%%%%%%%%%%%%%%%%%%%%%%
\section{Conclusion}

In this paper, we  have discussed a geometric picture of the
separable set of states for a general pure bipartite state,
 based on algebraic complex projective geometry. In
particular, we have proved that complete separability  for a
general pure bipartite state can be seen as a Segre variety.
 Moreover, we have generalized this result to multi-partite states, by
 defining a map called multi-projective Segre embedding. The image
 of this map defines a quadric space, namely the generalized Segre
 variety which we constructed by a prime ideal of two-by-two subdeterminants of a
 so-called decomposable tensor. We showed that the Segre variety define
 the completely
 separable states of a general multipartite state.
  Furthermore, based on this
 subdeterminant, we define an entanglement measure for general
 pure bipartite  and three-partite states which coincide with generalized
 concurrence.

\begin{flushleft}
\textbf{Acknowledgments:} The authors acknowledge useful
discussions with Professor Ingemar Bengtsson. This work was
supported by the Swedish Research Council (VR), the Swedish
Foundation for Strategic Research (SSF), and the Wenner-Gren
Foundation.
\end{flushleft}

%%%%%%%%%%%%%%%%%%%%%%%%%%%%%%%%%%%%%%%%%%%%%%%%%%%%%%%%%%%%%%%%%%

\end{document}